\documentstyle[prl,aps,epsfig,graphicx,twocolumn]{revtex}

\begin{document}

\title{\bf Simulation of polysilane and polysilyne formation and 
structure}

\author{R.L.C. Vink} 
\address{
	Instituut Fysische Informatica, 
	Utrecht University, 
	Leuvenlaan 4, 
	3508 TD Utrecht, the Netherlands }

\author{G.T. Barkema} 
\address{
	Theoretical Physics, 
	Utrecht University, 
	Princetonplein 5, 
	3584 CC Utrecht, the Netherlands }

\author{C.A. van Walree and L.W. Jenneskens} 
\address{
	Debye Institute, Department of Physical Organic Chemistry, 
	Utrecht University, 
	Padualaan 8, 
	3584 CH Utrecht, the Netherlands }

\date{\today}

\maketitle

\begin{abstract}

We present Monte Carlo simulations of the formation and structure of
polysilane, hybrid polysilane/polysilyne and polysilyne networks. The
simulation technique allows for the investigation of large networks,
containing up to 1000 silicon atoms. Our results show that ring formation
is an important factor for all three types of materials. For polysilyne
networks, a random structure is found incorporating cyclic substructures,
linear chains and branching points.

\end{abstract}

\section{Introduction}

The electronic and optical properties of silicon-based materials are
strongly related to the structure of the silicon skeleton and the size and
surface properties of the materials~\cite{brus}. Crystalline silicon,
which consists of a three-dimensional silicon framework, is an indirect
semi-conductor with a band gap of around 1.10 eV. In other forms of
three-dimensional silicon, such as nano-crystalline and porous silicon the
band gap is somewhat larger and quasi-indirect, so that rather efficient
photo- and electroluminescence have been observed. In contrast,
one-dimensional polysilanes \mbox{(-SiR$_2$-)$_n$}, which are linear
polymers consisting of a backbone of silicon atoms to which two organic
side groups R are bonded~\cite{west,miller} exhibit well-defined, intense
absorption and emission bands in the UV spectral region. An interesting
aspect of the electronic spectra of polysilanes is that both in the solid
state and in solution they are often strongly temperature dependent. This
thermochromism finds its origin in conformational changes of the silicon
backbone, which is said to be $\sigma$-conjugated. Polysilanes are usually
synthesized by polymerization of SiR$_2$Cl$_2$ monomers, employing alkali
metals as coupling agent. The heterogeneous polymerization process is
rather complicated, but is proposed to involve both silyl radicals and
silyl anions as reactive intermediates~\cite{jones}. It is noteworthy that
in the alkali metal-mediated polymerization there is a tendency to form
cyclic oligosilane oligomers.

Organic side chain appended silicon-based materials with a higher
dimensionality than polysilanes can be obtained by polymerization of
trichlorosilanes RSiCl$_3$~\cite{bianconi1,bianconi2}. This yields
so-called polysilynes, in which each silicon atom is bonded to three other
silicon atoms and to one organic side group R. Polysilynes exhibit a broad
indirect semiconductor-like absorption which tails into the visible
region~\cite{bianconi2,furukawa,wilson1}. The fluorescence, emanating from
trapped excitons, is situated in the visible region and has a broad
appearance. Hybrid polysilane/polysilyne networks, obtained by
polymerization of a mixture of dichloro- and trichlorosilanes, have also
been reported~\cite{walree,sasaki,wilson2,watanabe}. These kind of
polymers can be viewed as linear polysilanes in which SiR branching points
are incorporated. In hybrid polysilane/polysilyne networks the
photophysical properties of polysilanes and polysilynes are more or less
combined.

Since their discovery, the structure of polysilynes and hybrid
polysilane/polysilyne networks has been the subject of debate. Polysilynes
were initially assumed to consist of a rigid network of interconnected
ring-like structures~\cite{bianconi2}, but it was also argued that they
form essentially two-dimensional sheetlike networks~\cite{furukawa}.  
According to another point of view it was reasoned that the growth of a
polysilyne preferably occurs at the termini of the polymer chain, which
leads to a hyperbranched, dendritic morphology~\cite{maxka}. Cleij {\it et
al.} used PM3 calculations on silyl radicals and silyl anions to show that
chain propagation preferentially occurs at the termini of oligo- and
polymers and is more likely to occur than branching~\cite{cleij}. It
furthermore appeared that formation of silicon ring structures is also an
aspect to take into account. Hence, the PM3 calculations led to the idea
that polysilynes possess a predominant one-dimensional structure with
small branches and incorporated rings.

A number of experimental observations also indicates that polysilynes and
hybrid polysilanes/polysilynes can be viewed as essentially linear
structures. For a series of $n$-hexyl substituted hybrid
polysilyne/polysilane networks thermochromism (both in the solid state and
in solution), fluorescence and a degree of exciton delocalization were
observed which resembled the properties of linear
polysilanes~\cite{walree}. Even more surprising, for a polysilyne with
oligo(oxyethylene) side chains in aqueous environment thermoresponsive
behavior very similar to that of linear polysilanes was
found~\cite{cleij}. This thermoresponsive behavior has to originate from
folding and unfolding processes of the silicon framework. These results
imply that the silicon backbone of polysilynes and hybrid
polysilyne/polysilanes are to a certain extent flexible and behave much
like one-dimensional systems. Similar conclusions can be drawn from the
photophysical properties of well-defined oligosilane
dendrimers~\cite{lambert,suzuki}.

In this contribution, we present computer simulations of the formation of
polysilane, hybrid polysilane-polysilyne networks and polysilynes, and
study the properties of the resulting networks. There are three essential
characteristics that we included in these simulations: 1) the starting
point is a random mixture without any polymerization; 2) monomers diffuse,
and when they meet they can form stable bonds; 3) conglomerations of
bonded monomers are not stationary and rigid, but they show some diffusion
and flexibility.  More specifically, in our simulations we implemented the
time evolution by means of Monte Carlo dynamics (since the dynamics is
overdamped because of the solvent), and the elastic properties of
conglomerations were described by an empirical interaction potential
featuring bond-stretching and bond-bending, with parameters chosen to
match experimental properties of crystalline Si.  It is anticipated that a
thorough knowledge of the structure of polysilanes and polysilynes gives
more insight in the properties of these silicon-based materials.

\section{Model of polysilane and polysilyne}

\begin{figure} 
\begin{center}
\epsfxsize=5cm
\epsfbox{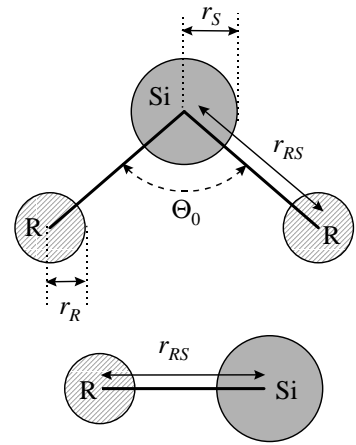}
\end{center} 
\caption{Diagrammatic representation of SiR$_2$ (top) and SiR fragments
(bottom).  Silicon atoms are labeled 'Si', alkyl groups 'R'. Their respective
radii are $r_S$ and $r_R$. The ground state Si-R distance is set $r_{RS}$ and
the ground state R-Si-R angle in silane to the tetrahedral angle $\Theta_0$.}
\label{fig:molecules} 
\end{figure}

The simulations are started without any trace of polymer present. The only
bonds initially present in the system are therefore the internal Si-R
bonds shown in Fig.~\ref{fig:molecules}. However, in the course of the
simulation, as monomers react with each other, Si-Si bonds also form. In
these simulations, both the silicon atoms and the alkyl groups are treated
as hard spheres with radii $r_S$ and $r_R$, respectively, see
Fig.~\ref{fig:molecules}. In the ground state, the Si-R distance is set to
$r_{RS}$ and the R-Si-R bond angle to the tetrahedral angle $\Theta_0$,
with $\cos(\Theta_0)=-1/3$. During the simulation, as the monomers diffuse
and react, distances and angles are allowed to fluctuate around their
ground state values.

In the model, all non-bonded particles interact according to hard-sphere
potentials. Bonded particles interact according to the Keating
potential~\cite{keating}, which describes the stiffness of the network with
respect to bond-length and bond-angle distortions.  This potential requires an
explicit list of all bonds, and is given by
\begin{eqnarray} 
V &=& \frac{3\alpha}{16} \sum_{\langle ij \rangle} \frac{1}{d_{ij}^2} 
\left( {\bf r}_{ij} \cdot {\bf r}_{ij} - d_{ij}^2 \right)^2 \nonumber \\ 
&&+ \frac{3\beta}{8} \sum_{\langle jik \rangle} \frac{1}{ d_{ij} d_{ik} } 
\left( 
{\bf r}_{ij} \cdot {\bf r}_{ik} + \frac{1}{3} d_{ij} d_{ik} \right)^2, 
\end{eqnarray} 
where the summation runs over all pairs and triples in the system; $\alpha$
and $\beta$ are the bond-stretching and bond-bending force constants,
respectively;  ${\bf r}_{ij}$ is the vector pointing from particle $i$ to
particle $j$ and $d_{ij}$ is the ground-state distance between particles $i$
and $j$; if these are both silicon atoms, $d_{ij}$ equals $r_{SS}$, otherwise
one particle is a silicon atom and the other an alkyl fragment, and $d_{ij}$
equals the equilibrium Si-R distance $r_{RS}$ in that case. The values for the
parameters used in the simulation are listed in Table~\ref{table:param}.

\section{simulation dynamics} 
\label{subsec:simul}

The simulation starts with a configuration of $n_2$ SiR$_2$ and $n_3$ SiR
monomers, placed at random positions and with random orientations in a
periodic box, under the constraint of the hard-sphere interactions.  The
size of the box is chosen such that the particle density equals $2.0 \cdot
10^{-4}$ monomers per \AA$^3$, matching the experimental conditions where
typically 20 mmol of monomer is put into a reaction volume of 60
ml~\cite{walree}. The nature of the dynamics is two-fold: monomer
diffusion and reactions between monomers. We simulate this with the
approach of rare-event dynamics~\cite{red}. In this approach, the two
types of events are monomer displacements, which occur with a rate $r_1$,
or bond formation between two silicon atoms, which occur with a rate
$r_2$. This reaction can only occur when (1) both Si atoms are not
saturated, i.e., they have less than four bonds to either alkyl-groups or
other silicon atoms, (2) the silicon atoms are separated less than a
cut-off distance $r_c$, and (3) the silicon atoms are not mutually bonded
already.  We continue to simulate until reactions become rare.  This
occurs when approximately 80\% of all possible Si-Si bonds have formed. We
then investigate the structure of the final network.

When concentrating on the final structure, the overall scaling of the
rates $r_1$ and $r_2$ is irrelevant, since it only affects the time scale
of the simulation.  The networks presented here were generated using
$r_1:r_2=5:1$.  To the best of our knowledge, no experimental data on the
reaction rates $r_1$ and $r_2$ are reported.  However, we have established
that the networks are insensitive to the ratio of rates. For $r_1:r_2=1:2$
and $1:10$, we obtained similar results.

We recall that polysilanes are formed by silicon atoms bonded to two alkyl
groups, and polysilynes by silicon atoms bonded to one alkyl group.  
Therefore, a system consisting of $n_2$ SiR$_2$ and $n_3$ SiR monomers is
represented by a total number of $N \equiv 3n_2+2n_3$ locations in
three-dimensional space. We denote the total number of pairs of silicon
atoms that are able to react as $P$, a quantity that varies during the
simulation. In terms of $N$ and $P$, the total diffusion rate $R_1$ can be
written as $R_1=r_1 N$, the total reaction rate $R_2$ as $R_2=r_2 P$, and
the total rate of events as $R_1+R_2$. Events are selected one-at-a-time.
To each event, a time-increment of $\Delta t=1/(R_1+R_2)$ is attributed.
The likelihood that this event is a diffusion event or a reaction event,
is equal to $R_1\Delta t$ and $R_2 \Delta t$, respectively.

To describe the diffusion process, we introduce the following event:
\begin{enumerate} 
\item We select randomly one silicon atom or alkyl group.
\item For this silicon atom or alkyl group, a displacement is proposed, 
drawn randomly from within a sphere with radius $r_m$.
\item If the hard-sphere constraints are violated, the proposed displacement
is rejected. Otherwise, the displacement is accepted with the Metropolis
acceptance probability~\cite{metrop}
\begin{equation} 
P_m = \min \left[1, \exp \left( \frac{E_b-E_f}{k_b T} \right) \right], 
\end{equation} 
where $k_b$ is the Boltzmann constant, $T$ is the temperature, and $E_b$ 
and $E_f$ are the total (Keating) energies of the system before and after 
the random displacement. 
\end{enumerate} 
Since we displace single silicon atoms and alkyl groups, the above
procedure also allows for the vibration and rotation of monomers.

To describe reactions in the simulation, we introduce the following 
reaction event: 
\begin{enumerate} 
\item We select randomly one of the $P$ pairs of silicon atoms 
able to react. 
\item A bond is placed between the two silicon atoms constituting the 
selected pair. 
\end{enumerate}

\section{Results}

Polysilane networks, polysilyne networks and hybrid polysilane-polysilyne
networks are obtained using the simulation method described above. Each
system contains a total of 1000 monomers.  To characterize the structure
of these networks, we proceed as follows. First, we identify clusters,
defined as a group of connected monomers. For each cluster, we calculate
its genus $g$ given by $g=1+e-n$, where $e$ is the number of Si-Si bonds
in the cluster and $n$ the number of silicon atoms in the cluster. The
genus helps to identify the topology of the cluster: it measures the
number of bonds that can be cut before the cluster loses its connectivity.  
For example, if $g=0$, the cluster is a chain of connected monomers,
i.e.~a polymer, and if $g=1$, the cluster is a ring, possibly with a
number of side-chains attached. Networks with high values of $g$ have a
complicated topology.

\subsection{Polysilane}

\begin{figure} 
\begin{center}
\epsfxsize=8cm
\epsfbox{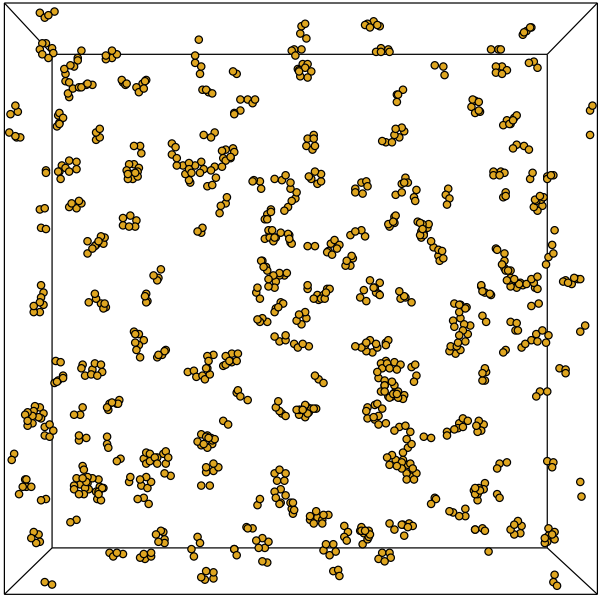}
\end{center} 
\caption{Resulting polysilane structure at density $2.0 \cdot 10^{-4}$
monomers per \AA$^3$. For clarity, the alkyl groups are not shown.}
\label{fig:silane_lowconc} 
\end{figure}

We performed one simulation containing 1000 SiR$_2$ and no SiR fragments
at a density of $2.0 \cdot 10^{-4}$ monomers per \AA$^3$. The final
network is shown in Fig.~\ref{fig:silane_lowconc}; 86.9\% of all possible
Si-Si bonds are formed. 

\begin{table} 
\caption{Values of parameters used in the simulation: $r_{RS}$ and
$r_{SS}$ are the ground state Si-R and Si-Si distances, respectively, and
$r_S$ and $r_R$ are the radii of a silicon atom and an alkyl group,
respectively; values for these four parameters were adapted from an
MM2-calculated structure of dodecamethycyclohexasilane. Reactions between
two silicon atoms can occur only when their separation is less than $r_c$.
$r_m$ is the maximum particle displacement during a diffusion step.
$\alpha$ and $\beta$ are the standard Keating
parameters~\protect\cite{keating}. $T$ is the temperature; $r_1$ and $r_2$
are the rates of diffusion and reaction, respectively.}
\begin{tabular}{lll} 
$r_{RS}$	& 1.89   & \AA 	\\
$r_{SS}$	& 2.35   & \AA 	\\
$r_S$		& 1.20   & \AA 	\\	
$r_R$		& 1.10   & \AA 	\\
$r_c$		& 2.50   & \AA 	\\
$r_m$		& 0.30   & \AA 	\\
$\alpha$	& 2.965  & eV \AA$^{-2}$   \\	
$\beta$		& 0.845  & eV \AA$^{-2}$  \\
$T$		& 293.0  & K 	  \\    	  			 
$r_1:r_2$	& $5:1$  & 	  \\	                           	     
\end{tabular} 
\label{table:param} 
\end{table} 

\begin{figure} 
\begin{center}
\epsfxsize=8cm  
\epsfbox{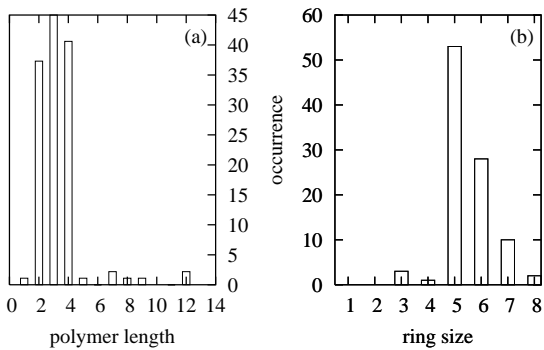}
\end{center} 
\caption{Histograms showing polymer length (a) and ring size distribution
(b) for polysilane consisting of 1000 monomers at density $2.0 \cdot
10^{-4}$ fragments per \AA$^3$.}
\label{fig:silane_lc} 
\end{figure}

Since silane is bifunctional, only two types of structure exist in this
case: chains (polymers) with $g=0$ and rings with $g=1$. The network
consists of 228 separate clusters, ranging in size from one up to twelve
silicon atoms. Of these clusters, 131 are chains, the remainder are rings.
Polymer length and ring size statistics are shown in
Fig.~\ref{fig:silane_lc}. In our simulation, 46.8\% of the silane monomers
form linear structures; a substantial fraction of 53.2\% of the monomers
is found in rings. For this system, the Si-R and the Si-Si bond lengths
are $2.10\pm0.07$ \AA\ and $2.32\pm0.08$ \AA. The R-Si-R and Si-Si-Si
bond angles are $107.0\pm5.1$ and $109.7\pm7.9$ degrees, respectively.

\begin{figure} 
\begin{center}
\epsfxsize=8cm  
\epsfbox{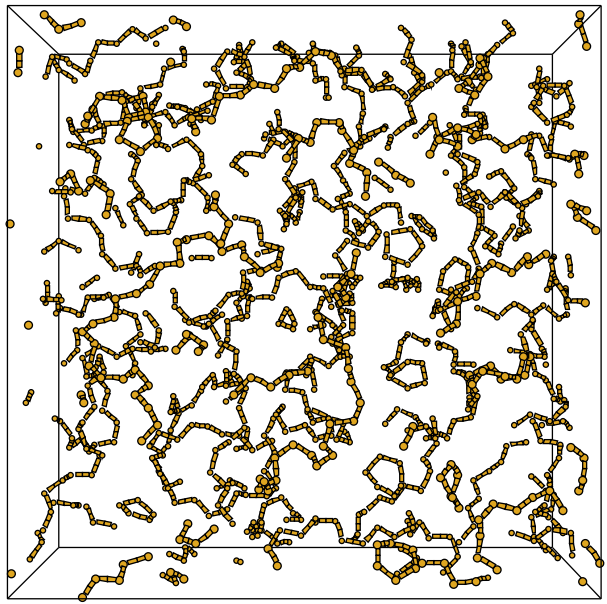}
\end{center}
\caption{Resulting polysilane structure at density $5.0 \cdot 10^{-3}$
monomers per \AA$^3$. For clarity, the alkyl groups are not shown.}
\label{fig:silane_highconc} 
\end{figure}

\begin{figure} 
\begin{center} 
\epsfxsize=8cm  
\epsfbox{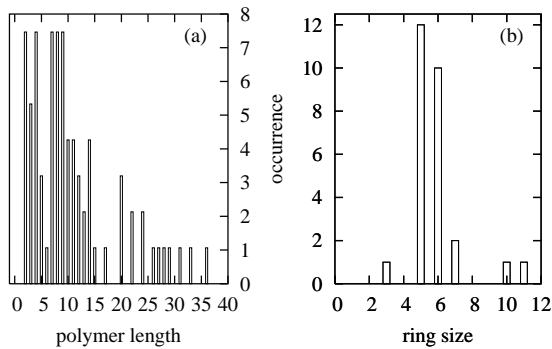}
\end{center} 
\caption{Histograms showing polymer length (a) and ring size statistics
(b) for polysilane consisting of 1000 monomers at density $5.0 \cdot
10^{-3}$ monomers per \AA$^3$.}
\label{fig:silane_hc} 
\end{figure}

To study the effect of the monomer density, an additional polysilane is
simulated at an increased density of $5.0 \cdot 10^{-3}$ monomers per
\AA$^3$. This network is shown in Fig.~\ref{fig:silane_highconc}; 92.3\%
of all Si-Si bonds are formed.  In this case, the system contains 104
separate clusters, ranging in size from two up to 36 silicon atoms. Of
these clusters, 77 are incorporated in linear structures, the remainder
form rings.  Fig.~\ref{fig:silane_hc} shows the polymer length and ring
size distributions. In this simulation, 84.2\% of the silane fragments
form linear structures; the remainder form rings. For this system, the
Si-R and the Si-Si bond lengths are $2.09\pm0.07$ \AA\ and $2.32\pm0.08$
\AA. The R-Si-R and Si-Si-Si bond angles are $107.2\pm5.0$ and
$111.1\pm6.7$ degrees, respectively.

\subsection{Polysilyne}

\begin{figure} 
\begin{center}
\epsfxsize=8cm  
\epsfbox{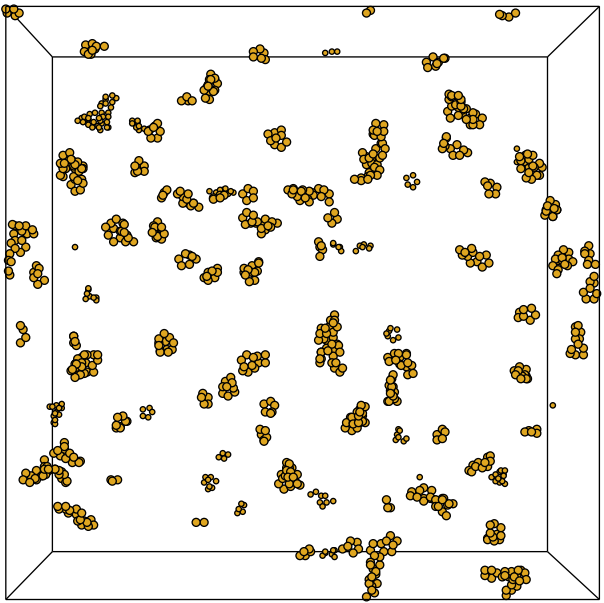}
\end{center} 
\caption{Polysilyne network at a density of $2.0 \cdot 10^{-4}$ monomers 
per \AA$^3$. For clarity, the alkyl groups are not shown.}
\label{fig:silyne_lowconc} 
\end{figure}

\begin{figure} 
\begin{center} 
\epsfxsize=8cm  
\epsfbox{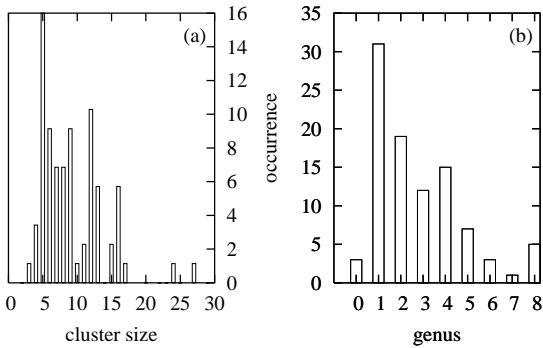}
\end{center} 
\caption{Histogram of cluster size (a) and genus number (b) of a silyne
network consisting of 1000 monomers at density $2.0 \cdot 10^{-4}$ 
monomers per \AA$^3$.}
\label{fig:silyne_lc} 
\end{figure}

We also performed one simulation involving 1000 SiR monomers, no SiR$_2$
monomers being present at a density of $2.0 \cdot 10^{-4}$ monomer
fragments per \AA$^3$.  The final network is shown in
Fig.~\ref{fig:silyne_lowconc}; 77.9\% of all possible Si-Si bonds were
formed. Since silyne is trifunctional, the genus of the structures can now
reach large values. The final network consists of 96 separate clusters,
ranging in size from two up to 29 monomers. Fig.~\ref{fig:silyne_lc} shows
a histogram of the cluster sizes. Of these clusters, only three are
chains, 31 possess a monocyclic structure and the remaining are structures
with $g>1$. Fig~\ref{fig:silyne_lc}b shows a histogram of the genus
numbers. The structure is not dendritic: by far the largest fraction of
monomers are part of rings; the fraction of monomers in chains or
side-chains is 5.6\% only. Moreover, the chains and side-chains are short,
consisting at most of three monomers. For this network, the Si-R and the
Si-Si bond lengths are $2.09\pm0.08$ \AA\ and $2.33\pm0.09$ \AA,
respectively. The Si-Si-Si bond angle is $106.7\pm10.0$ degrees.

\begin{figure} 
\begin{center} 
\epsfxsize=8cm 
\epsfbox{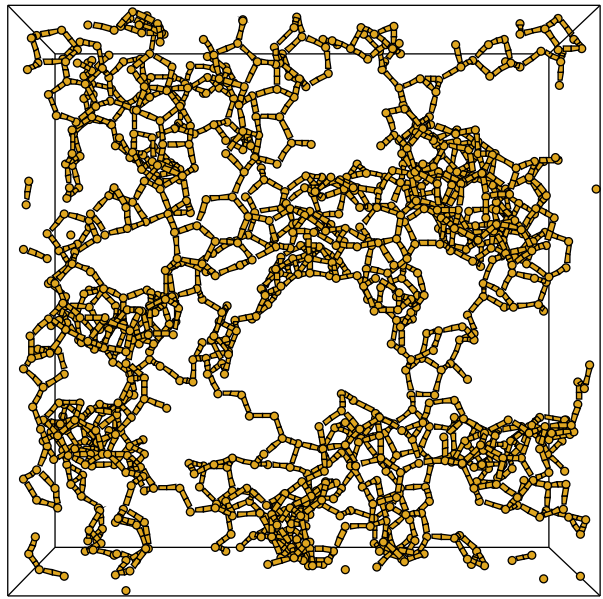} 
\end{center} 
\caption{Polysilyne network at a density of $8.0 \cdot 10^{-3}$ monomers
per \AA$^3$. For clarity, the alkyl groups are not shown.}
\label{fig:silyne_highconc} 
\end{figure}

\begin{figure} 
\begin{center}
\epsfxsize=8cm  
\epsfbox{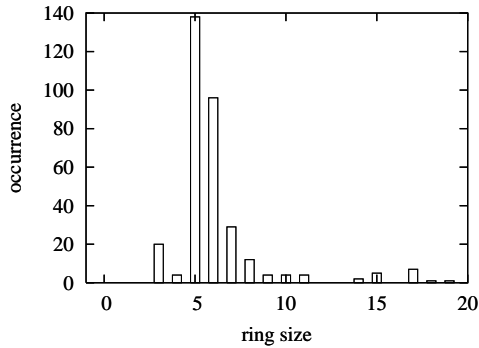}
\end{center} 
\caption{Ring size histogram for a polysilyne cluster consisting of 1000
monomers. This cluster was generated at a density of $8.0 \cdot 10^{-3}$
monomers per \AA$^3$.}
\label{fig:silyne_hc} 
\end{figure}

To study the effect of the monomer density, an additional silyne network
is generated at an increased density of $8.0 \cdot 10^{-3}$ monomers per
\AA$^3$. This network is shown in Fig.~\ref{fig:silyne_highconc}. In this
case, 88.8\% of all possible Si-Si bonds are formed. All monomers form one
large cluster with a genus of 333. The structure consists mostly of
connected rings; 94.4\% of the monomers are part of rings. The ring
statistics are shown in Fig.~\ref{fig:silyne_hc}.

\subsection{Hybrid silane/silyne networks}

\begin{figure} 
\begin{center}
\epsfxsize=8cm  
\epsfbox{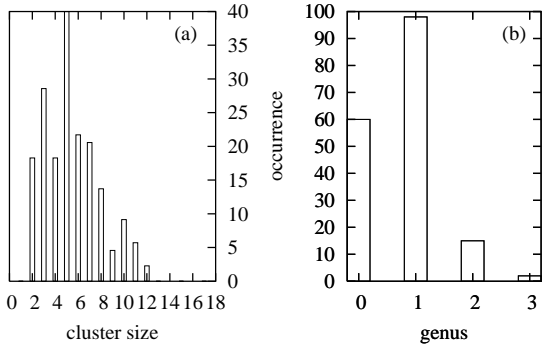}
\end{center} 
\caption{Histogram of cluster size (a) and genus number (b) for a 
silane-rich network, consisting of 800 silane and 200 silyne 
monomers.} 
\label{fig:silane-rich} 
\end{figure}

We also performed simulations of hybrid silane/silyne networks, either
rich in silylene units (containing 800 SiR$_2$ and 200 SiR monomers) or
rich in branching points (containing 200 SiR$_2$ and 800 SiR monomers).  
The density was $2.0 \cdot 10^{-4}$ monomers per \AA$^3$. For both
networks, approximately 80\% of the possible Si-Si bonds are formed. The
SiR$_2$-rich network consists of 175 separate clusters, ranging in size
from one up to 17 monomers. In Fig.~\ref{fig:silane-rich}a, a histogram of
the cluster sizes is shown. Of these clusters, 60 are linear structures,
98 are rings, the rest are structures with $g>1$. See
Fig.~\ref{fig:silane-rich}b for a histogram of the genus numbers found.  
For this network, the typical chain length ranges from two to four
monomers; rings consist typically of five to six silicon atoms.

\begin{figure} 
\begin{center}
\epsfxsize=8cm  
\epsfbox{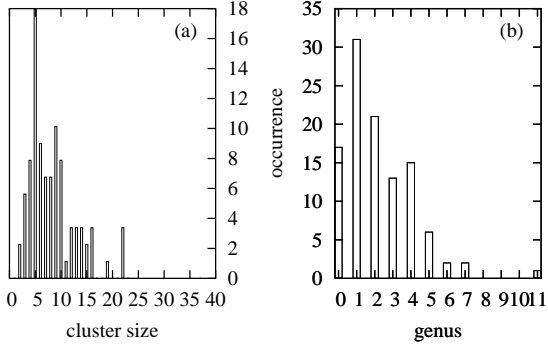} 
\end{center} 
\caption{Histogram of cluster size (a) and genus number (b) for a 
silyne-rich network, consisting of 200 silane and 800 silyne 
monomers.} 
\label{fig:silyne-rich} 
\end{figure}

The silyne-rich network consists of 108 separate clusters, ranging in size
from one up to 37 monomers. In Fig.~\ref{fig:silyne-rich}a, we show a
histogram of the cluster sizes found. Of these clusters, 17 are linear
structures, 31 are rings, the rest are structures with $g>1$. See
Fig.~\ref{fig:silyne-rich}b for a histogram of the genus numbers found.
For this network, the typical chain length is two monomers; rings
typically contain five to six silicon atoms.

Both networks show structures that are not dendritic. For the silane-rich
network, 31.8\% of the monomers form chains. For the silyne-rich network,
this fraction is 11.6\%. Most of the monomers are incorporated in rings.

\section{Discussion and Conclusion}

The Wurtz-type polymerization of dichlorodialkylsilanes and
trichloroalkylsilanes is a complex process~\cite{west,miller,jones}. The
polymerization proceeds by a chain growth process at the alkali metal
surface, which plays an essential role in the formation of polymers of
high molecular weight. This is for instance indicated by the fact that
employment of homogeneous coupling agents leads to formation of only
oligomeric materials~\cite{miller,zeigler}. Which products actually are
obtained also depends on parameters as the nature of the alkali metal, the
solvent and the type of organic side chain. Moreover, after the initial
stages of the formation of linear polymers, secondary reactions including
depolymerization, backbiting by silyl anions and redistribution of chain
lengths are known to occur~\cite{west,miller,jones}. For polysilynes,
secondary reactions such as ring opening reactions (by analogy with
spirosilanes, see Ref.~\cite{maxka}), ring-size redistribution and
coupling to linear moieties may be of interest. It is a virtually
impossible task to perform a simulation of a Wurtz-type polymerization at
a metal surface, hereby also taking all these contributing factors into
account.  Nevertheless, as the approach we used led to results which
adequately reflect experimental findings, we believe that our results are
sound and provide insight into the structure and properties of
polysilanes, polysilynes and hybrid polysilane/polysilyne networks.

For polysilane, for instance, the simulations adequately predict the
effect of the monomer density on the nature of the silicon compounds
formed. At low density, the system has a strong tendency to form five- and
six-membered rings, see Fig.~\ref{fig:silane_lc}. This finding is in
accordance with the experimental observation that formation of cyclic
silicon-based compounds is favored by slow addition of the monomer to the
alkali metal, i.e.~by maintaining a low monomer
concentration~\cite{brough,chen}. The ring size distribution shown in
Fig.~\ref{fig:silane_lc}, which indicates that five-membered rings are
formed in excess, is in agreement with the experimental distribution
such as obtained upon reaction of a dialkyldichlorosilane under kinetic
control~\cite{west}. It is noteworthy that in an equilibrium distribution,
which arises from redistribution of the kinetic mixture upon use of excess
alkali metal, the six-membered ring dominates. As expected, the amount of
long linear fragments increases strongly when the monomer density is
increased, see Fig.~\ref{fig:silane_hc}. At this higher concentration
five- and six-membered rings are formed in equal amounts. Furthermore, the
length of the linear fragments is rather moderate, which may support the
idea that the presence of an alkali metal surface is essential for the
formation of high molecular weight polymers~\cite{miller,zeigler}.

For the polysilyne, the simulations also predict the formation of cyclic
structures, see Figs.~\ref{fig:silyne_lowconc}-\ref{fig:silyne_hc}. At low
concentration virtually all monomers are incorporated into rings.
Notwithstanding that monocyclic systems are formed in the largest amount,
the number of mutually bonded rings is already considerable. At high
monomer density, a very complex network of genus 333 is obtained. Of the
monomers, 94.4\% is incorporated in rings, which mainly contain five or
six silicon atoms. Thus, whereas in the linear case a higher concentration
leads to a larger amount of linear structures, for the polysilynes the
silicon atoms are still virtually exclusively found in cyclic structures.
A few linear fragments are however also present. The simulated structure
is in close agreement with the characteristics of the material that was
isolated after the first stage of a polymerization of
trichloroalkylsilanes~\cite{uhlig}. Our simulations indicate that the
structure of polysilynes can be viewed as a network of silicon atoms
comprising fused cyclic structures and cyclic structures mutually
connected by a single Si-Si bond. There is neither indication of formation
of a regular sheetlike arrangement of silicon atoms~\cite{furukawa} nor of
a dendritic structure~\cite{maxka}, at least in our simulations where
there is no difference in reactivity between Si-SiCl-Si groups and
terminal SiCl groups.

The results obtained for the hybrid polysilane / polysilyne networks do
not substantially differ from those obtained for the linear or fully
branched systems. With 20\% SiR monomers present, mainly short linear and
monocyclic compounds are formed, see Fig.~\ref{fig:silane-rich}. However,
the average cluster size is larger than in the linear case, and already
some clusters of genus two and three, i.e.~bicyclic and tricyclic systems
are found. When the number of branching points is increased to 80\%, the
average cluster size and the genus increase considerably. This
demonstrates once again that incorporation of trifunctional (SiR) monomers
leads to a structure with interconnected and fused rings. Hence, the
results for the hybrid polysilane and polysilyne networks show that this
class of materials is built up from cyclic structures interconnected via
short linear chains.

An interesting question is to what extent the structure of the polysilyne
depicted in Fig.~\ref{fig:silyne_highconc} is consistent with the
flexibility of polysilynes such as indicated by its thermal properties,
i.e.~whether the structure allows for conformational changes affecting the
degree of $\sigma$-conjugation. Although at first sight the network of
fused cyclic structures seems to be quite rigid, there are a number of
features which suggest a certain degree of flexibility. Firstly, some
(short) linear fragments and extended cyclic structures, both of which
imply conformational flexibility, are distinguishable in
Fig.~\ref{fig:silyne_highconc}. Secondly, there is also a number of rings
which are interconnected by a single Si-Si bond. This should give the
possibility for rings to rotate with respect to each other; i.e., the
structure depicted in Fig.~\ref{fig:silyne_highconc} is not rigid. A third
factor which may contribute to the flexibility of the networks is that
silicon rings are known to be highly flexible. In this context it is of
interest that even in the solid state dodecamethylcyclohexasilane
Si$_6$Me$_12$ undergoes a rapid ring inversion~\cite{casarini}. While it
is not likely that ring inversions occur for fused ring systems, they
might be possible for rings connected to linear chains or connected to
other rings by a single Si-Si bond.

Hence, the networks obtained in our simulations are already flexible.  
However, there are a number of reasons why experimentally prepared
polysilynes can even more easily undergo conformational changes. A
preferred polymer extension at the termini of the already formed polymer,
as discussed above, will enhance chain growth rather than branching. This
will have the consequence that polysilynes in reality may contain more
linear chains and larger cyclic structures than depicted in
Fig.~\ref{fig:silyne_highconc}. Another factor is the size of the organic
side group R. While in the simulations R has the dimensions of a methyl
group, much more bulky side groups such as hexyl, isobutyl and phenyl
substituents have been used in experimental
studies~\cite{walree,furukawa}. If bulky side groups are present the
network is expected to be less dense and to incorporate more linear
fragments than in the present case.

A question left to be answered is whether polysilynes obtained by a
Wurtz-type condensation should be regarded as one, two or
three-dimensional silicon materials. At first sight, the silicon network
in Fig.~\ref{fig:silyne_highconc} percolates in three dimensions. However,
the description of the simulated polysilyne and hybrid
polysilane/polysilyne networks by $g = 1+e-n$, which is Euler's equation
for two-dimensional networks~\cite{devlin}, implies that topologically the
silicon backbones of these materials are best regarded as two-dimensional
systems. This is consistent with the electronic spectra which approach
that of an indirect band gap
semiconductor~\cite{bianconi2,furukawa,wilson1}.

In summary, we have shown that the formation and structure of polysilanes,
polysilynes and hybrid-polysilane/polysilyne networks can be adequately
described by a simulation based on rare event dynamics of diffusion and
reaction steps. The simulations indicate that ring formation is an
important factor for all three types of silicon materials. Insight has
been obtained in the conformational flexibility of polysilynes and
polysilane/polysilyne materials such as inferred from experimental
studies. It is anticipated that simulations can provide an even more
comprehensive picture when variation of the size of the organic
substituent R and differences in reactivity of different types of Si-Cl
functionalities are implemented. This would be the subject of further
work.

\end{document}